# Noise disturbance and lack of privacy: Modeling acoustic dissatisfaction in open-plan offices

Manuj Yadav  ; Jungsoo Kim  ; Valtteri Hongisto  ; Densil Cabrera  ; Richard de Dear 

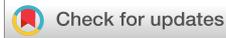

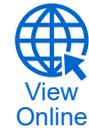 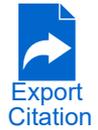

View Online  Export Citation

---

### Articles You May Be Interested In

Tuning the cognitive environment: Sound masking with "natural" sounds in open-plan offices

*J. Acoust. Soc. Am.* (April 2015)

Subjective and objective assessment of acoustical and overall environmental quality in secondary school classrooms

*J. Acoust. Soc. Am.* (January 2008)

Comparison of two speech privacy measurements, articulation index (AI) and speech privacy noise isolation class (NIC′), in open workplaces

*J. Acoust. Soc. Am.* (May 2002)

# Noise disturbance and lack of privacy: Modeling acoustic dissatisfaction in open-plan offices

Manuj Yadav,[1,2,a)] Jungsoo Kim,[1] Valtteri Hongisto,[3] Densil Cabrera,[1] and Richard de Dear[1]

[1]School of Architecture, Design and Planning, The University of Sydney, New South Wales, 2006, Australia
[2]Institute for Hearing Technology and Acoustics, RWTH Aachen University, Kopernikusstrasse 5, Aachen 52074, Germany
[3]Turku University of Applied Sciences, Built Environment, Joukahaisenkatu 3-5, FI-20520 Turku, Finland

**ABSTRACT:**
Open-plan offices are well-known to be adversely affected by acoustic issues. This study aims to model acoustic dissatisfaction using measurements of room acoustics and sound environment during occupancy, and occupant surveys (n = 349) in 28 offices representing a diverse range of workplace parameters. As latent factors, the contribution of *lack of privacy* (LackPriv) was 25% higher than *noise disturbance* in predicting *acoustic dissatisfaction* (AcDsat). Room acoustic metrics based on sound pressure level (SPL) decay of speech ($L_{p,A,s,4m}$ and $r_C$) were better in predicting these factors than distraction distance ($r_D$) based on speech transmission index. This contradicts previous findings, which may partly be due to the cross-sectional study design. Specifically, the trends for SPL-based metrics in predicting AcDsat and LackPriv go against expectations based on ISO 3382-3. For sound during occupation, $L_{A,90}$ and psychoacoustic loudness ($N_{90}$) predicted AcDsat, and a SPL fluctuation metric ($M_{A,eq}$) predicted LackPriv. However, these metrics were weaker predictors than ISO 3382-3 metrics. Medium-sized offices exhibited higher dissatisfaction than larger (≥50 occupants) offices. Dissatisfaction varied substantially across parameters including ceiling heights, number of workstations, and years of work, but not between offices with fixed seating compared to more flexible and activity-based working configurations.

## I. INTRODUCTION

### A. Noise disturbance and lack of privacy in open-plan offices

Since the early days of open-plan offices (OPOs), it was already suggested that the "coexistence of good acoustics and open planning" may be difficult to achieve.[1] Over the years, the acoustic issues in OPOs have been linked primarily to speech permeating across workstations from neighboring conversations, lobbies, meeting areas, etc.[2–6] Further, there is evidence that non-speech sounds, e.g., telephone rings, walking sounds, can be disruptive too.[4] A recent large-scale survey reported acoustic issues—distracting speech/noise and lack of speech privacy—as the most common source of dissatisfaction among the indoor environmental quality (IEQ) factors in offices (62 360 respondents).[7] Another survey involving 82 315 respondents reported noise being the most common source of environmental dissatisfaction in flexible offices.[8] These findings further crystallize overwhelming evidence about acoustic issues being some of the most problematic aspects in OPOs. Such evidence has been summarized in studies in each decade[9–15] since the popularization of OPOs in the 1960s.[16] Moreover, acoustic issues have largely persisted regardless of the design philosophy including landscaped[10] and cubicle-based designs[17] to more recent designs such as flexible offices[8] and those configured to Activity-Based Working (ABW).[18]

At the core of these issues is the conflict between communication and privacy requirements of the occupants.[14,17] The intrusion of irrelevant speech/sounds across workstations can cause disruptions especially during cognitively demanding tasks, i.e., noise disturbance, and create the perception that one's own private conversations will likely be heard, i.e., lack of speech privacy. In OPOs, visual privacy is typically compromised as well, which, along with the inability to regulate one's working environment,[19] compounds the general sense of lack of privacy.

This leads to the hypothesis that overall *acoustic dissatisfaction* can be explained by *noise disturbance* and *lack of privacy* (see Fig. 1). The interrelationship between these factors has been largely putative in previous studies, and the relative contribution of lack of privacy and noise disturbance towards acoustic dissatisfaction has not been studied directly. This is important from a design perspective as different measures may be relevant in addressing lack of privacy and noise disturbance, and in turn, acoustic dissatisfaction. Further, most extant studies refer to one or more of these factors using individual survey questions.[7,8] This precludes a more detailed characterization of these factors, which would instead require using a larger set of

---

[a)]Email: manuj.yadav@sydney.edu.au

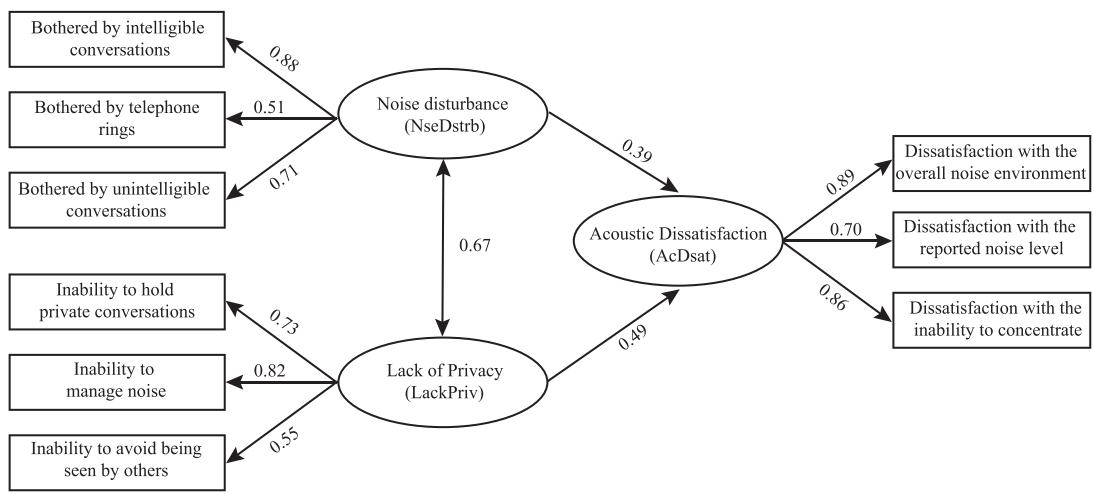

FIG. 1. Structural equation modeling output showing the Factors (latent variables; as ovals), Survey items (indicator variables; rectangles) that manifest as ratings on these factors, factor loadings between the factors and manifest variables, correlation coefficient between noise disturbance and lack of privacy, and the regression coefficients predicting acoustic dissatisfaction.

survey questions that refer to underlying aspects. While the study by Pierrette et al.[4] used a questionnaire addressing satisfaction with various IEQ aspects in OPOs with a particular focus on acoustics [subsequently included within ISO 22955[20] (Annex D)], their analysis was mostly descriptive and did not examine the intercorrelations between survey items or systematically explore the factors they may relate to.

### B. Instrumental characterization of OPO acoustics

To complement survey findings, instrumental acoustic characterization of OPOs can broadly involve measurements in *unoccupied* (i.e., room acoustics) and/or *occupied* (i.e., sound environment during working hours) conditions. Room acoustic research for OPOs has a rich tradition, with the reduction of speech intelligibility, or in basic terms speech-to-noise ratio, being the cornerstone;[1,21–23] see summary in Virjonen et al.[24] ISO 3382-3[25] continues this tradition, with the key idea being to characterize room acoustic quality based on speech decay along one or more paths of workstations in unoccupied OPOs.[25,26] This is different to previous approaches that involved measurements between two adjacent workstations[27] and/or global metrics including reverberation time.[28]

The ISO 3382-3 metrics include those based on spatial decay of A-weighted speech sound pressure level (SPL) and speech transmission index (STI) from the sound source. They include A-weighted speech SPL at 4 m distance ($L_{p,A,s,4m}$, in dB), spatial decay rate of A-weighted speech SPL with distance doubling ($D_{2,S}$, in dB), comfort distance ($r_C$, in m)—distance where A-weighted speech SPL falls below 45 dB, and distraction distance ($r_D$, in m)—distance where STI falls below 0.5, besides the average A-weighted SPL of background noise ($L_{p,A,B}$) over workstations.

Measurements in occupied conditions are inherently non-trivial as they can be affected by the occupancy rates, measurement durations, etc.[29] To account for some such aspects, ISO 22955[20] attempts to consolidate several previous approaches, where the emphasis is on speech reduction across workstations. ISO 22955 includes SPL-based room acoustic metrics from ISO 3382-3 ($D_{2,S}$ and $L_{p,A,s,4m}$), and the more traditional reverberation time ($T_r$ in s). It further includes metrics that attempt to characterize the sound environment in occupied offices: sound at workstations measured as A-weighted equivalent continuous SPL over time $T$ ($L_{A,eq,T}$ in dB), and A-weighted speech attenuation at a measurement point relative to 1 m in free field ($D_{A,S}$ in dB).[20] Unlike ISO 3382-3, ISO 22955 does not consider background noise and speech intelligibility (e.g., distraction distance) in unoccupied conditions. This omission is notable given that most previous OPO room acoustic research focuses on speech intelligibility.[1,22]

Crucially, most metrics and guidelines, including those in these two standards, have not been tested extensively in real office environments. The main exception is $L_{A,eq}$ due to office activities, which has been the most widely used metric to characterize occupants' noise ratings in offices across studies spanning over 50 years,[29] although not always successfully.[30] Further, for sound measurements during occupation, there is wide variation in methods and uncertainties across studies.[29] In terms of investigating the relationship between room acoustic quality described by the ISO 3382-3 metrics and subjective perception of noise, Haapakangas et al.[5] is the main empirical study thus far. The cross-sectional study design in Haapakangas et al. included convenience sampling of 17 OPOs (21 in total, 4 excluded in the final model) measured over a roughly 12-year period. They reported $r_D$ as the most consistent predictor of occupants' noise- and speech-based disturbance: increased $r_D$ values were associated with increased probability of high noise disturbances.[5] This is consistent with Hongisto et al.,[31] who reported that refurbishments in a call centre OPO that improved ISO 3382-3:2012 metrics ($L_{p,A,s,4m}$ and $r_D$ but not $D_{2,S}$) were associated with reduced distraction due to many contributors to environmental noise (including nearby speech) and improvements in acoustic privacy. In Park et al.,[6] however, none of the ISO 3382-3 metrics predicted noise disturbance. Instead, $L_{A,eq,8h}$ at workstations in occupied OPOs predicted noise disturbance.[6]

The inconsistent results of Park et al.[6] and Haapakangas et al.[5] could partly be due to the smaller sample size in the former comprising seven OPOs with distinct room acoustics. Moreover, some of the $r_D$ values in Haapakangas et al. may vary by more than 2.3 m due to inconsistencies and some sources of uncertainty in the STI measurement methods used compared to the ISO 3382-3 method.[32] Further, the variation in high noise disturbance was extremely large in Haapakangas et al. at certain $r_D$ values: noise disturbance varied from 10 to 60% when $r_D$ was within 8–10 m. Collectively, this demonstrates a well-known limitation of cross-sectional studies: the results can vary across different samples.[33]

In addition to the metrics mentioned above, other relevant metrics used previously to predict occupants' perceptions of workplace noise include percentile levels (e.g., $L_{A10}$, $L_{A50}$;[30] part of ISO 22955:Annex E[20]) and SPL fluctuation relative to the background noise level during occupancy (typically $L_{A90}$), e.g., noise climate (NCl; $L_{A10}$-$L_{A90}$),[34] and $M_{eq}$ ($L_{A,eq}$-$L_{A90}$).[35] Another possibility includes psycho-acoustic parameters for characterizing sound fluctuation (e.g., fluctuation strength), and other aspects such as loudness, sharpness, etc., which have rarely been considered in OPO studies.[29]

In general, the variety of room acoustics[36] and the range of activity noise levels[29] is extremely large in OPOs. While some prediction models for noise/speech disturbance exist,[5,6] the prediction of lack of privacy and overall acoustic dissatisfaction using metrics characterizing unoccupied/occupied OPOs has rarely been the focus of previous studies. More work is needed to investigate occupants' perception of acoustic issues in OPOs within large-scale studies similar to Haapakangas et al.,[5] although with a consistent measurement method. There is also a need to represent divers workplace parameters (e.g., room acoustics, work type, occupant density, etc.) to mitigate the weakness inherent within cross-sectional studies, and include measurements of both room acoustics and sound during occupation. This paper presents such an investigation.

### C. Aims of the study

There are two main aims. First, the hypothesis that overall acoustic dissatisfaction can be explained by noise disturbance and lack of privacy (see Sec. I A and Fig. 1) is tested using structural equation modeling. An adapted version of the questionnaire in ISO 22955[20] will be used for this purpose. Second, acoustic dissatisfaction, noise disturbance, and lack of privacy will be modeled using metrics based on instrumental acoustic measurements in OPOs, and several workplace and personnel-related parameters.

## II. METHODS

### A. General information about the study

Data acquisition was conducted over a two-year period (2017–2018) and involved conducting acoustic measurements (Sec. II B) and administering an occupant survey (Sec. II C) in OPOs. The sample here includes **n = 28** OPOs that were located in nine buildings in metropolitan Australia. The sampling strategy included contacting several building managers, property owners, and acoustic consultants about the study's aims (i.e., convenience sampling). Most OPOs were mixed functions to an extent, e.g., administration offices had some customer service staff. Yet, the primary work activities in these OPOs could broadly be categorized as seen in Table I.

Each OPO represented an acoustic zone with a consistent absorption profile, room height, and workstation layout as described in ISO 3382-3.[25] While all the OPOs had centralized heating, ventilation, and air-conditioning (HVAC) systems, none had additional electroacoustic masking systems. The surface area (Table I) included the portion of floor plate dedicated to just the working areas, excluding areas for building services (e.g., elevator lobby), kitchens, enclosed rooms (e.g., for meetings). This may partly account for the relativelysmaller areas in Table I compared to some previous studies.[6]

As seen in Table I, most OPOs were carpeted and had sound absorptive flat ceilings, although some had complicated ceiling designs. Almost two-thirds of the OPOs had no partitions between workstations other than a computer screen. The remaining OPOs included 1-, 2-, and/or 3-sided partitions (latter being a cubicle) ranging in heights from 1.1 to 1.6 m. Just under a half of the OPOs included ABW with no pre-allocated seating for workers, who were instead encouraged to choose a workspace from either OPO areas with workstations (usually no partitions), enclosed meeting rooms, collaboration areas, etc. to suit their activity. For OPOs with ABW, it was nevertheless quite common for teams to occupy a certain area of workstations regularly,

TABLE I. Summary of key workplace parameters from 28 offices in nine buildings. $M$: Mean, Standard deviation: $SD$.

| Parameter | Summary |
| --- | --- |
| Number of workstations | $M$ ($SD$), Range: 39 (17), 16 – 78 |
| Workstation density (per 100 m$^2$) | $M$ ($SD$), Range: 12.6 (6.0), 4 – 24 |
| Ceiling height (m) | $M$ ($SD$), Range: 3.2 (1.3), 2.7 – 7.6 |
| Surface area (m$^2$)[a] | $M$ ($SD$), Range: 221.3 (153.3), 80.5 – 719 |
| Ceiling type | Absorptive = 21 (75%), Hard = 7 (25%) |
| Carpet | Yes = 22 (78.6%), No = 6 (21.4%) |
| Activity-based working | Yes = 13 (46.4%), No = 15 (53.6%) |
| Partition | Yes = 10 (35.7%), No = 18 (64.3%) |
| Work activities | A = 4(14.3%), B = 7(25.0%), C = 3(10.7%), D = 13(46.4%), E = 1(3.6%)[b] |

[a]Indicates just the working areas.
[b]A: Architecture, Design; B: Adminstration; C: Engineering; D: Management; E: Customer Service.

TABLE II. Metrics of unoccupied OPOs (n = 28).

| Metric | Unit | M (SD) | Range |
|---|---|---|---|
| $T_{30}$ | s | 0.6 (0.3) | 0.3–1.2 |
| $L_{p,A,B}$ | dB | 42.5 (4.2) | 35.6–51.0 |
| $D_{2,S}$ | dB | 5.1 (1.0) | 2.7–7.4 |
| $L_{p,A,s,4m}$ | dB | 52.1 (2.3) | 46.1–54.9 |
| $r_C$ | m | 12.1 (5.1) | 4.6–30.2 |
| $r_D$ | m | 10.4 (2.5) | 4.4–17.0 |

and conversations at or between workstations within the OPO areas were also quite common.

### B. Instrumental acoustic measurements

ISO 3382-3 compliant room acoustic measurements were conducted in 36 OPOs with normal HVAC operation during unoccupied hours; detailed description is provided in a previous publication that uses the current data.[37] Out of these, data from n = 28 OPOs is included here, summarized in Table II (further details in the supplementary material), where $T_{30}$ and $L_{p,A,B}$ are averaged values per workstation path. For the remaining metrics, since some workstation paths were measured with the loudspeaker at both ends,[37] the respective values in Table II are averaged values per path.

Measurements of sound during working hours was conducted in 43 OPOs.[29] Out of these n = 28 OPOs are included in this study where the room acoustic measurements were also possible. The measurements included a binaural dummy head (Neumann KU100; Berlin, Germany) placed at an ear height of 1.2 m representing a seated listener, and at least 0.5 m away from the desk. These measurements were performed at several vacant workstations or nearby locations to cover the entire OPO as best as possible (≥30% of the workstations). The overall sampling strategy was to approximate typical sound environment for workers. Each measurement was at least four hours long (meeting recommended duration in ISO 22955: Annex E[25]) and some OPOs were measured for up to a week.[29] Occupancy during the measurements fluctuated but was generally over 80%. Table III provides key SPL-based (including those in ISO 22955) and psychoacoustic metrics derived from these measurements. The latter included binaural loudness (N), sharpness (S), roughness (R), and fluctuation strength (FS).[38] For all metrics besides N, the value reported is the average value of the two ears averaged over all measurements per OPO. The highest $L_{p,A,B}$ value is much higher than the highest $L_{A,90,4h}$ value, which was due to differences in the measurement locations in some offices during room acoustic and occupied sound measurements. These offices are still included here as their removal did not change the results substantially. More details including calculation of psychoacoustic metrics, etc. are provided in a previous publication.[29]

TABLE III. Metrics characterizing occupied OPOs.

| Metric | M (SD) | Range |
|---|---|---|
| $L_{A,eq,4h}$ (dB) | 53.9 (2.9) | 48.3–58.5 |
| $L_{A,10,4h}$ (dB) | 57.4 (3.1) | 51.6–62.5 |
| $L_{A,50,4h}$ (dB) | 47.5 (3.4) | 42.4–53.3 |
| $L_{A,90,4h}$ (dB) | 32.6 (3.5) | 27.6–38.7 |
| $NCI$ (dB) | 24.8 (1.5) | 22.6–30.2 |
| $M_{eq}$ (dB) | 21.3 (1.8) | 18.4–27.6 |
| $N_{90}$ (sone) | 4.6 (0.9) | 3.2–6.3 |
| $S_{mean}$ (acum) | 1.2 (0.1) | 1.0–1.4 |
| $FS$ (vacil) | 0.3 (0.1) | 0.1–0.6 |
| $R_{max}$ (asper) | 4.1 (1.6) | 1.2–7.1 |

### C. Occupant survey

An online survey was administered to OPO occupants in Australia. The survey was adapted from a previous study[4] which was subsequently included as Annex D in ISO 22955.[25] The current survey used questions that were rated on continuous semantic differential scales (SDS) with bipolar adjectives at the extremes per scale. The survey (see the supplementary material) was divided into five sections of which four sections are relevant here: general information about the occupant; level of agreement about dissatisfaction with various IEQ factors (SDSs with "Totally" and "Not at all" at the extreme); multiple choice questions, and questions about various aspects of the sound environment at the workstations (SDSs; "Not at all" – "Totally"); and level of agreement with various statements about noise sensitivity (SDSs; "Completely agree" – "Completely disagree"). Each SDS had an underlying scale of 0–100, and ratings were made by moving a horizontal slider.

The researchers had no direct contact with the occupants. The online survey was made accessible via a link sent to the occupants by the building manager per OPO. Participation was voluntary and personal information was anonymized. Occupants were allowed to fill in the survey over an approximately two-week period, which generally coincided with acoustic measurements. To minimize biases due to extreme conditions (e.g., too quiet/loud), the occupants were asked to give their "long-term" assessments. Each office had at least 10% of the occupants completing the entire survey (*Mean* = 35.1%, *Range* = 11.1%–63.3%), and included 349 occupants in total [Age (years): *Mean (SD)* = 38.4 (10.5), *Range* = 21–80; *females* = 55.6%; participants with *Fixed desks* = 53.8%].

### D. Statistical analysis

All statistical analyses were conducted within the software R. To characterize survey responses in terms of the hypothesis in Sec. I C, the strategy included starting with an exploratory factor analysis (Sec. II D 1) followed by full structural equation modeling (SEM) (Sec. II D 2). The factors resulting from the SEM analysis are then modeled against metrics (Sec. II B) using Bayesian regression in Sec. II D 3.

*1. Exploratory factor analysis*

To understand the underlying factor structure and to reduce the number of survey items related to sound perception, exploratory factor analysis (EFA) was conducted using the R package *psych*[39] (version 2.2.9). Items with a Kaiser-Meyer-Olkin (KMO) measure greater than 0.7 were kept for further analysis based on the criterion that they contributed "moderately" to the common variance. Following this, the overall KMO was 0.9, which represents "excellent/marvelous" suitability of the sample size for factor analysis.[40] Bartlett's test for sphericity [$\chi^2(45) = 1973, p < 0.001$] indicated that the correlation matrix was suitable for EFA. Three factors were retained based on the inspection of the scree plot and parallel analysis.

The factor loadings were determined using maximum likelihood (ML) estimation and *oblimin* rotation. Using the common cut-offs for fit indices,[41] the final model showed a good fit with low values for the residual statistics: root mean square error of approximation (RMSEA) = 0.06, standardized root mean square of the residuals (SRMR) = 0.03; and high values for goodness-of-fit statistics: Tucker-Lewis Index (TLI) = 0.97, Comparative Fit Index (CFI) = 0.99. A criterion of at least 0.4 loading was used to ensure that the survey item was related to underlying factor. Items that loaded on more than one factor were excluded.

Based on the EFA, the three items loading onto Factor 1 included those relating to the level of dissatisfaction with the *overall noise environment* and *inability to concentrate*, and whether the *level of noise in the work area was high*. Factor 1 is hence considered to represent the **Acoustic Dissatisfaction (AcDsat)**. The three items loading onto Factor 2 asked how bothered the participants were with the *intelligible* and *unintelligible* conversations, and *telephone rings* around their workstation. Factor 2 hence represents **Noise Disturbance (NseDstrb)**. The three items loading onto Factor 3 asked about the level of dissatisfaction with the inability to hold *private conversations*, manage *noise*, and to *avoid being seen by others* at their workstation (i.e., visual privacy). Factor 3 is considered to represent **Lack of Privacy (LackPriv)**.

*2. Structural equation modeling*

The three-factor model from EFA was used to test the hypothesis (Sec. IC) that AcDsat can be explained by NseDstrb and LackPriv. This was done using structural equation modeling (SEM) using the R package *lavaan*[42] (version 0.6–13). Figure 1 presents the coefficients based on ML estimation of the final model, which had a good fit (CFI = 0.99, TLI = 0.99, RMSEA = 0.04, SRMR = 0.03) based on the common cut-off criteria for fit indices,[41] except for a significant $\chi^2$ value [$\chi^2(24) = 38.3, p < 0.05$]. The latter, however, is sensitive to sample sizes. Hence, overall the model fit is considered good based on other indices.

*3. Bayesian mixed-effects modeling*

For subsequent analyses, the estimated value of each manifest variable on its latent factor, i.e., factor scores, from the SEM models were used (Sec. II D 2). These were calculated using the Bartlett approach (*lavPredict* function in *lavaan*), which produces estimates most likely to represent "true" factor scores.[43] The factor scores were dichotomized by median splitting the scores, where scores less than the median represented occupants that were "highly dissatisfied" (HD) with that factor. This is similar to Haapakangas *et al.*[5] who had used HD to refer to "highly disturbed" with noise and speech.

The %HD occupants per office for each factor (dependent variable) was then modeled using each of the acoustic and psychoacoustic metrics (Tables II and III) as the independent variables in separate Bayesian mixed-effects regression models. The *%HD occupants per office* for each factor will be simply referred to using the factor's label (i.e., AcDsat, NseDstrb, and LackPriv) in the following. The random effect of buildings (where the measurements were conducted) was incorporated as independently varying intercepts. The Bayesian modeling was done using the *brms* (version 2.19) package[44] with weakly informative priors. Prior distributions were sampled 40 000 times: 4 independent chains of 11 000 samples each, and discarding the first 1000 warm-up samples. The calculated posterior probability distributions (PPDs) are summarized using the median values and associated 95% Bayesian credible intervals (CIs). The latter were calculated using the highest density interval of the PPD rather than the quantiles to provide a more comprehensive uncertainty assessment. Given the model, the 95% CI (different from the 95% confidence interval in frequentist statistics) is the interval with a 95% chance of containing the effects' true value. Statistically robust effects refer to the 95% CI not spanning zero. Bayesian $R^2$ values are used to compare performance across predictors. Marginal $R^2$ values are reported in the following, which describe the variance accounted for by the predictor (fixed-effect) alone, and not the random effects. The commonly used Cohen's criterion[45] is used to interpret the effect size of $R^2$ values.

## III. RESULTS AND DISCUSSION

### A. Structural equation modeling results

Building upon previous studies,[14] the model in Fig. 1 provides a comprehensive account of the dichotomous role of communication (NseDstrb) and lack of privacy (LackPriv) aspects in characterizing acoustic dissatisfaction (AcDsat) in offices. The regression/path coefficients in Fig. 1 show that LackPriv accounts for AcDsat around 25% more than NseDstrb. This may partly be due to LackPriv encompassing a diverse concept, which is indicated here by ratings for speech privacy, visual privacy, and noise management (highest loading onto LackPriv). NseDstrb instead is a more focused/singular concept, indicated here by high loadings for both intelligible and unintelligible conversations, besides telephone rings (lowest factor loading onto NseDstrb).

Compared to previous studies,[7] the model in Fig. 1 provides a more direct assessment of the relative contribution of the two commonly reported causes of AcDsat, and how the latter manifests in ratings of the overall noise environment, noise level, and the inability to concentrate. The terms speech/noise disturbance and privacy are sometimes used interchangeably and/or as referring to the same underlying latent factor.[13] However, the findings here (Fig. 1) stress that, while correlated, these factors encompass different qualities and responses from office workers, and would need to be addressed specifically within office design (discussed further in Sec. III B).

### B. Bayesian regression modeling using ISO 3382-3 metrics

For context, room acoustic improvements due to increased absorption and/or partitions, and/or furniture layout typically lead to increased $D_{2,S}$, and decreased $L_{p,A,s,4m}$ and $r_C$ (not accounting for the effect of room geometry). $r_D$ further incorporates HVAC noise and additional sound masking, with improvements in these aspects leading to reduced $r_D$ values. Based on this and some previous studies, Annex C of ISO 3382-3 provides values per metric indicating "good" and "bad" acoustic conditions. The tacit expectation here is that values within the good range should lead to better assessments from occupants, and *vice versa*.

Figure 2(A) shows that $L_{p,A,s,4m}$ and $r_C$ are statistically robust predictors of all the three factors outlined in the SEM analysis (Fig. 1). Since $r_C$ is calculated using $L_{p,A,s,4m}$ (and $D_{2,S}$), similar results for these metrics are not surprising. The $R^2$ values show that $L_{p,A,s,4m}$ accounts for a reasonably high proportion of variance per factor (with weak–substantial effect sizes), including the highest for AcDsat, whereas $r_C$ has relatively lower $R^2$ for this factor but relatively high $R^2$ values for the other two factors (moderate–substantial effect sizes). $D_{2,S}$ and $r_D$ are robust predictors, respectively, of only AcDsat (moderate effect size) and LackPriv (weak effect size). $L_{p,A,B}$ is a robust predictor of LackPriv and NseDstrb, with a moderately high $R^2$ value for the former. $T_{30}$, which is included in ISO 22955 but not ISO 3382-3, was not a robust predictor of any factors. Broadly, this indicates that the SPL-based metrics outperform $r_D$ as predictors of occupants' perception. However, this is discussed further in the following, where the focus is mostly on moderate–substantial effect sizes.

#### 1. Lack of privacy (LackPriv)

The negative slopes for $L_{p,A,s,4m}$, and $r_C$ in Fig. 2(A) indicate that as these values increase, LackPriv reduces overall, which is opposite to what would be expected. However, increasing $r_D$ increases LackPriv, which is as expected. Increasing $L_{p,A,B}$ (which indicates increasing sound masking) reduced LackPriv, which is expected to an extent although exorbitant $L_{p,A,B}$ values can be problematic. To contextualize these findings, we note first that the current sample is marked by around 64% of the offices without screens/partitions (Table I). This suggests limited sound absorption and visual privacy. Further, Figs. 3(A) and 3(B) show that some offices with high $r_C$ (low absorption), high $r_D$ (high intelligibility), and $L_{p,A,B}$ are not rated as unfavorably as would be expected based on Annex C. It is hence likely that insufficient sound absorption in some offices (i.e., seemingly adverse ISO 3382-3 metric values) may have been counterbalanced by the existence of, or even preference/tolerance for, relatively high HVAC noise levels. The latter may have contributed towards sound masking (i.e., higher sound privacy) for occupants from nearby sources in some cases. However, negative effects of Lombard speech (i.e., louder speech) are likely to be exhibited for $L_{p,A,B}$ exceeding 43.3 dB.[46] Hence, inordinately high HVAC noise level is not recommended here over sensible room acoustic planning, which would instead involve designing for sufficient level of sound masking along with sound absorption. The discussion above also underscores considering both SPL- and intelligibility-based metrics in unison. Further, the current findings stress that the assessment of lack of privacy (and room acoustics) is more involved than a rather simple distillation of good vs bad acoustic conditions as per Annex C, which was not particularly consistent in predicting LackPriv. The impact of LackPriv on AcDsat is particularly important given the model in Fig. 1 (see also Sec. III A) and needs to be specifically elaborated within ISO 3382-3 and ISO 22955.

In addition to the survey responses used for modeling, other questions included occupants being asked about their general approach(es) for noise management. When getting bothered by the noise at their workstation, 55% of the occupants chose "listening to music on headphones," 40% chose "taking a break," 33.5% "relocated elsewhere," and 7% mentioned strategies including "earplugs," "accepting lower productivity," "asking colleagues to respect quiet areas." This latter group of strategies emphasize the importance of office etiquette as mentioned in previous studies[20,22,47] A concerning aspect emergent is the use of headphones to listen to music, which has been shown to affect short-term memory performance[48,49] and is hence detrimental in general.

#### 2. Noise disturbance (NseDstrb)

As noticed in Fig. 2(A), $r_D$ is not a robust predictor of NseDstrb. Decreasing $L_{p,A,B}$ values show decreasing NseDstrb, which is as expected, although it is a weak effect. There is limited correspondence between $r_D$ and $L_{p,A,B}$ values and dissatisfaction rates [Fig. 3(C)]. The diminished role of $r_D$ is a major deviation from the main finding of Haapakangas *et al.*[5] (and ISO 3382-3 in general), where it was the most consistent predictor of speech and noise disturbance. The *M*, *SD*, and *Range* of $r_D$ values in Haapakangas *et al.* (excluding four offices with ABW) are 11.2, 3.2, and 4.8–18.0 m, respectively, which are similar to the current sample (Table II). However, $L_{p,A,B}$ values in their sample are quite low in general ($M = 35.5$ dB, $SD = 4.1$ dB, $Range = 29–43$ dB), and much lower than the current sample that has a higher mean and a wider range (Table II).

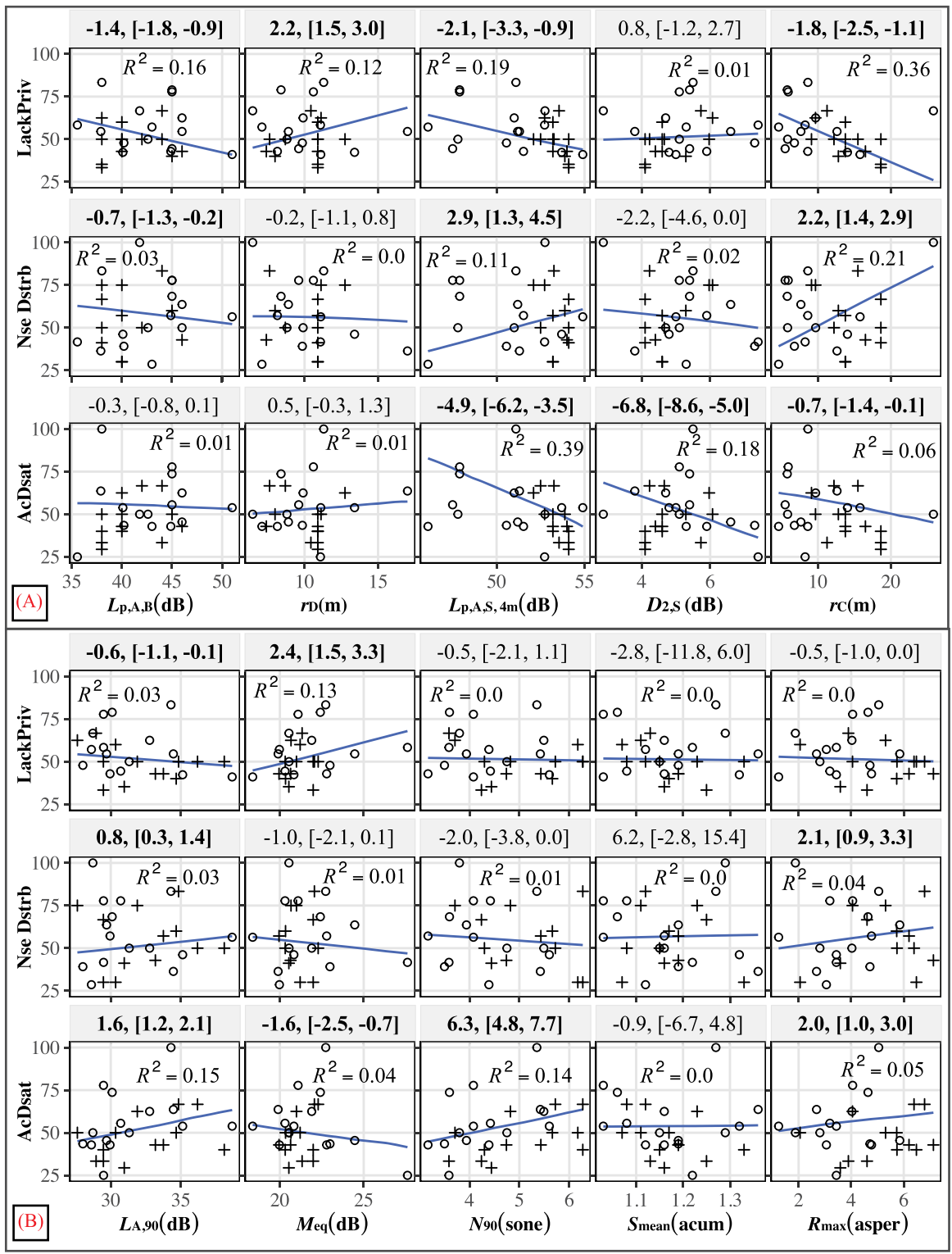

FIG. 2. Summary of Bayesian regression models predicting LackPriv, NseDstrb, and AcDsat using room acoustic (A) and occupied sound (B) metrics. Each scatterplot includes the median slope, whose value is presented in the shaded box above the scatterplot (e.g., 1 dB increase in $L_{p,A,B}$ reduces LackPriv by 1.4%) along with the 95% credible interval (CI); statistically robust values shown in bold. Offices that did and did not support ABW are plotted as (o) and (+), respectively. Marginal Bayesian $R^2$ values are provided. Cohen's criteria for interpretation of effect size: Ref. 45 $R^2 < 0.02$: Very weak, $0.02 \leq R > 0.13$: Weak, $0.13 \leq R^2 > 0.26$: Moderate, $R^2 \geq 0.26$: Substantial.

This is further illustrated in Fig. S1 in the supplementary material, where only one office in Haapakangas et al.[5] has $L_{p,A,B}$ within the good range in Annex C of ISO 3382-3. Hence, their sample included a high proportion of offices with relatively high sound absorption/screens, and relatively low levels of HVAC noise. It is then likely that their offices had inadequate speech/sound masking from nearby talkers (hence, diminished sound privacy as well), which is further evident from Fig. 3 in Haapakangas et al.[5] This may partly account for the more prominent role of $r_D$ in their study in predicting speech and noise disturbance over SPL-based metrics.

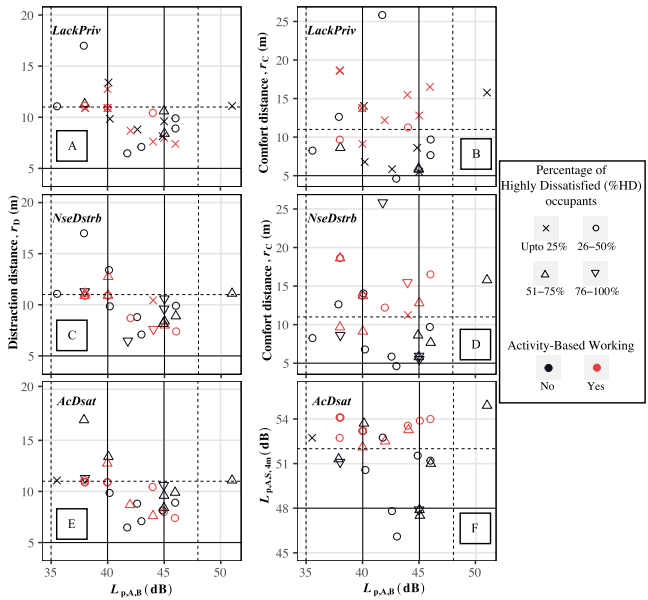

FIG. 3. Distribution of dissatisfaction rates in relation to key ISO 3382-3 metrics. As per Annex C in ISO 3382-3, good values indicated below the horizontal solid lines for the metrics on the $y$ axes ($L_{p,A,S,4\,m}$, $r_D$, and $r_C$), and bad values above the horizontal broken lines. Good values for $L_{p,A,B}$ are indicated in between the vertical solid lines, and bad values to the left and right of the dotted horizontal lines. Subfigures are labeled (A)–(F) for reference.

Instead, $L_{p,A,s,4m}$ and especially $r_C$ are better predictors of NseDstrb in the current study, with slopes in the expected directions. Most offices with high values of $r_C$ have higher dissatisfaction with NseDstrb [Fig. 3(D)]. These findings are inconsistent with Park et al.[6] since none of the ISO 3382-3 metrics were useful predictors in their models. $L_{p,A,s,4m}$ showed a trend similar to the current findings in Haapakangas et al.[5] when their complete sample was used but not after removing offices supporting ABW. In comparison, the current findings could be deemed more representative of modern workplaces by including an almost equal proportion of offices with and without ABW, besides a wide range of sound absorption and HVAC noise profiles (Table I).

The role of NseDstrb compared to LackPriv seems more consistent with the overall message of ISO 3382-3 in that increased absorption (reduced values of $L_{p,A,s,4m}$ and $r_C$) reduced dissatisfaction with noise disturbance in general, which is as expected. Note, however, that even exorbitant sound absorption alone is unlikely to be beneficial against noise from nearby sources.[50] Although HVAC noise is considered in $r_D$, which in principle is more informative than the speech SPL-based metrics, the findings here suggest that for offices over a wide range of $L_{p,A,B}$, the role of SPL-based metrics in ISO 3382-3 may be more important. This, however, needs to be tested over an even larger sample.

In addition to the responses used for modeling, the survey included a multiple choice question about speech disturbance. Therein, the majority (52%) of the occupants reported being the most bothered with intelligible multi-talker conversations (same percentage as in Pierrette et al.[4]) 26% when they could clearly hear only one talker, and 17.5% finding both single and multi-talker conversations equally disturbing. This finding supports a developing hypothesis that speech disturbance in OPOs is mainly due to multi-talker speech.[51] This is relevant for future investigations focusing on realistic simulations of OPO sound environments, and the potential use of "babble" speech (currently a rather vague term for a composite of multiple voices) in sound masking systems, etc.

### 3. Overall acoustic dissatisfaction

Figure 2(A) indicates that the association of AcDsat with $L_{p,A,B}$ and $r_D$ was not robust. Instead, the SPL-based sound decay metrics—$L_{p,A,s,4m}$, $D_{2,S}$, and $r_C$—are more useful in explaining AcDsat. However, the trend of this relationship is not as expected. In particular, a 1 dB increase in $L_{p,A,s,4m}$ is predicted to reduce AcDsat by roughly 5% (highest effect size among the SPL-based metrics). Decreasing sound absorption can potentially create conditions of lower intelligibility, with spectro-temporal blending of successive sounds in relatively strong reverberant sound fields (i.e., higher sound masking). However, such a monotonic trend would not be expected much beyond the range of values here, or assist cognitive performance in general in offices.[52] Further, unlike LackPriv (Sec. III B 1), the role of $L_{p,A,B}$ cannot be used here for explanations that combine balancing sound absorption and HVAC noise. One possibility is that AcDsat is a complex factor that cannot be explained primarily using ISO 3382-3 metrics, i.e., room acoustic characterization of unoccupied offices; Figs. 3(E) and 3(F) show no clear trends of dissatisfaction rates aligning with good room acoustic values. This, however, does not imply that improvements in ISO 3382-3 do not influence AcDsat. Figure 2 shows that an increase in SPL and loudness of the background sound during occupation is associated with increasing AcDsat, which is more in line with expectations. Further, these metrics of sound during occupancy increase with increasing values of $L_{p,A,s,4m}$ (i.e., decreasing sound absorption). For instance, a unit increase in the latter is associated with almost a unit increase in $L_{A,eq}$, which is a statistically robust effect (Bayesian slope = 1.01, 95% CI = [0.72, 1.32], $R^2 = 0.49$). This indicates that AcDsat may be better understood using assessments of the sound during occupancy, which incorporates occupants' perception of multi-talker speech and nonspeech sounds and other non-acoustic aspects that go beyond room acoustic measurements.

Moreover, an increase in AcDsat manifests as reported increase in noise levels, dissatisfaction with the overall noise environment, and increasingly difficult conditions for concentration (Fig. 1). The latter two are arguably aspects that may not be sufficiently represented by room acoustic quality assessments using ISO 3382-3. However, since AcDsat can be predicted using LackPriv and NseDstrb with the latter being closest in principle to the ISO 3382-3 philosophy, the metrics therein are necessary to address at least parts of the overall acoustic design for offices.

### C. Bayesian regression modeling using metrics of sound during occupancy

The metrics considered in this section represent the complex sound environment during working hours. Hence, their relationship with the occupants' perception arguably represents a closer assessment of the overall acoustic environment than the ISO 3382-3 metrics that characterize spatial decay due to speech from a single talker. However, unlike ISO 3382-3, these metrics cannot be incorporated directly into room acoustic design, and can vary depending on non-acoustic factors, e.g., day of the week. Although Yadav et al.[29] reported limited variation in SPL values over a week in several offices, more variation is possible in some offices.

Figure 2(B) shows that $L_{A,90,4h}$ is a robust predictor of all three factors in the SEM analysis ($L_{A,50,4h}$ as well). $L_{A,eq,4h}$ is a robust predictor of AcDsat only ($R^2 = 0.04$; weak effect). $L_{A,eq}$ was a significant predictor of noise disturbance in Park et al.,[6] but has not been a consistent predictor of the perception of overall sound environment across studies.[3] Moreover, as the number of workstations increased in the offices studied here, the change in $L_{A,eq}$ values was not significant.[29] $L_{A,eq}$ has historically been the most commonly used metric to study sound in offices,[29] and is even included as an acoustic indicator with suggested values for design of workplaces in ISO 22955 (without due evidence). However, the current study offers limited support in its role as a predictor of occupants' sound perception. Instead, $L_{A,90,4h}$, which represents the SPL exceeded 90% of the time, i.e., the background sound level during occupancy, offers a better alternative, at least for predicting AcDsat with a moderate effect size: a 3 dB increase in $L_{A,90,4h}$ (i.e., double the sound power) is associated with a change of roughly 5% in AcDsat [Fig. 2(B)].

In addition to absolute levels, fluctuations in sound level is another attribute that has been useful in previous studies to characterize noise disturbance (summarized elsewhere[29]) Of the ones studied here, $M_{A,eq}$, a metric quantifying $L_{A,eq,4h}$ fluctuations above the background sound during occupation ($L_{A,90,4h}$) performed the best, and is a robust predictor of AcDsat (weak-sized $R^2$) and LackPriv (moderate-sized $R^2$) [Fig. 2(B)]. In particular, increasing $M_{A,eq}$ values, which indicate increasing sound fluctuation, are associated with increasing LackPriv. Previous findings show that increasing number of workstations are associated with significantly decreasing $M_{A,eq}$ values, and significantly increasing $L_{A,90,4h}$ values.[29] Combined, this suggests that larger offices with more sound/speech sources are louder but may in turn provide more sound masking to assist with privacy in general (see also Sec. III D 1). Note, however, that the SPL-based fluctuation parameters were based on 4-h averages across offices and may not represent the effect of shorter-term fluctuations.

Among the psychoacoustic metrics, the loudness of background sound, as characterized by binaural loudness exceeded 90% of the time ($N_{90,4h}$, in sones), is a robust predictor of AcDsat, with higher $R^2$ [moderate sized effect; Fig. 2(B)] than $N_{5,4h}$ and $N_{mean,4h}$ (0.03 and 0.09, respectively). However, despite $N_{90,4h}$ being a better representation of loudness perception, its calculation is computationally expensive. Hence, the findings here suggest the use of $L_{A,90,4h}$ for characterizing AcDsat instead. In addition to loudness, the other key psychoacoustic parameters studied here included sharpness, fluctuation strength, and roughness. The former two were not robust predictors of any of the factors, whereas maximum roughness (perception of relatively fast amplitude modulations between 15 and 300 Hz) is a robust predictor [Fig. 2(B)] of NseDstrb and AcDsat. However, due to the weak effect sizes ($R^2$ values) for the latter two models, it is unclear whether there is much value in considering roughness as a predictor [Fig. 2(B)]. Further, while an increase in the number of workstations significantly increased the fluctuation strength in the offices studied here,[29] that effect was not relevant in terms of occupants' perceptions.

Overall, the findings here suggest that there is some benefit in considering metrics characterizing AcDsat based on sound level and psychoacoustic loudness, and LackPriv based on SPL fluctuations. Moreover, the relationships between metrics and ratings are generally in line with expectations, in that more adverse sound environments during occupancy lead to poorer ratings. However, the metrics here could not sufficiently account for all factors outlined in Fig. 1. At the very least, SPL-based metrics of sound during working hours can complement the ISO 3382-3 metrics, including when the relationships between the latter and occupants' ratings are not intuitive (e.g., Sec. III B). From another perspective, the findings here highlight that aspects such as concentration in actual offices may vary with overall loudness of the sound environment. While loudness/level is typically shown to be irrelevant for short-term memory in mostly laboratory-based cognitive psychology studies (see recent summaries elsewhere[47,52]), the current findings indicate that in actual offices, loudness may indeed play a role.

### D. Modeling using non-acoustic aspects

This section discusses the findings about key workplace parameters in Table I, and personnel-related parameters. Each parameter was used as independent variable in a Bayesian mixed-effects model as described in Sec. II D 3, and the results are summarized in Fig. 4.

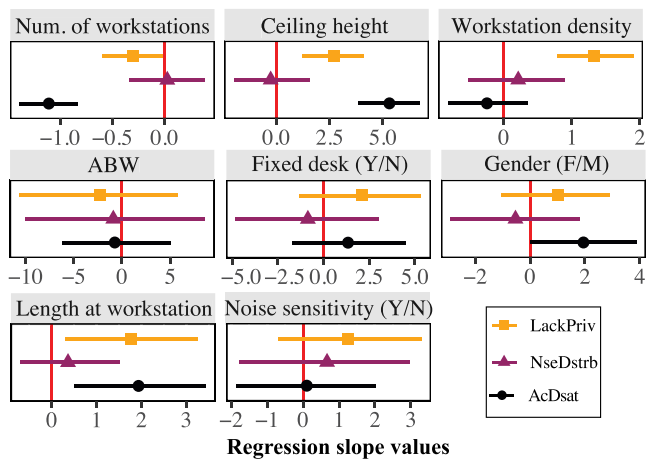

FIG. 4. Summary of models predicting LackPriv, NseDstrb, and AcDsat using workplace (top two rows) and (bottom two rows) personnel-related parameters. Median slope value (e.g., 1 m increase in ceiling height increased LackPriv by around 2.6%) and 95% credible intervals (CI) provided. Robust effects refer to the 95% CI not crossing zero. ABW (Activity-Based Working).

### 1. Workplace parameters

Figure 4 shows that LackPriv decreased with the number of workstations and increased with workstation density. AcDsat increased with increasing ceiling heights and decreased with increasing number of workstations (Table I). These trends are consistent with the findings in Sec. III C, where it was proposed that offices with more workstations potentially have more sound masking simply due to an increased number of sound sources. However, higher workstation densities can lead to higher LackPriv due to diminishing speech and visual privacy and the inability to manage noise from local sources. This is reinforced by considering the groups of offices with ≥50, 16–49, 7–15, and ≤6 occupants. Herein, medium-sized offices (7–15 people) had the highest LackPriv and AcDsat (statistically robust differences between groups). However, offices with 16–49 people (medium-large sized offices) had the highest NseDstrb, with statistically robust differences with offices with <50 and ≥50 occupants, with the latter having a lower ratings compared to offices with 7–15 occupants. Taken together, smaller offices seem to perform worse in occupant dissatisfaction. Moreover, number of workstations was the most important predictor of several metrics of sound during working hours in the offices studied here.[29]

However, an arbitrary increase in the number of workstations cannot be recommended based on these findings without considering the density. Yet, the metrics considered here may be interesting as indicators of the relationship of workplace size with occupant ratings, and assist in potential layout of offices, which was not directly addressed in this study. Other parameters including whether an office supported activity-based working or not, and whether occupants had fixed desks or not were not robust predictors of any factors (Fig. 4). This further suggests that the current sample and findings can be considered representative of a wide range of offices (e.g., those with/without ABW, and desk ownership).

### 2. Personnel-related parameters

Figure 4 shows that AcSat was around 2% higher for female occupants than male occupants. Each additional year at a fixed workstation increased LackPriv and AcDsat by roughly 2% as well. While both of these are robust effects, the gender difference is likely too small to be meaningful. Other categories including various age groups, and whether occupants were sensitive in general to noise or not were not robust predictors of any factors (Fig. 4).

### E. Limitations

The main limitation of this study is its cross-sectional design, which means that the results are likely to be dependent on the sample. The current sample does not include many offices that meet the criteria for good offices in Annex C of ISO 3382-3, and offices with electroacoustic sound masking. The sampling period for sound during occupancy was at least 4 h per location, which meets the ISO 22955 criterion. However, it is likely that longer duration is needed for a better characterization of sound during working hours. Further, while occupants were instructed to provide ratings based on long-term assessments, the influence of more temporary events in the offices cannot be ruled out. Necessary caution must be exercised as the results here may not be generalized much beyond the sample here. The questionnaire used here, while compliant with ISO 22955, does not represent the developing approaches in soundscape assessments,[53,54] and did not include questions about which could be addressed in future studies.

### IV. CONCLUSIONS

Overall, acoustic dissatisfaction is more strongly related to dissatisfaction with lack of privacy than noise disturbance. However, predicting these factors using existing metrics needs further work using even more diverse samples than here that combine instrumental measurements in both occupied and unoccupied offices, occupant surveys, and address key workplace parameters. The SPL-based metrics in ISO 3382-3 ($L_{p,A,s,4m}$ and $r_C$) were better predictors of AcDsat, LackPriv, and NseDstrb than $r_D$. This is inconsistent with previous studies. Moreover, the association of these SPL-based room acoustic metrics with occupants' ratings align with expected trends only for NseDstrb. For AcDsat and LackPriv, a more involved reasoning is required to explain the trends. This underlines the complexity in predicting occupants' ratings using objective metrics. Yet, the findings suggest that the criteria for good and bad acoustic conditions in ISO 3382-3 need a critical overhaul. Metrics characterizing sound in occupied offices are weaker predictors of occupant ratings than ISO 3382-3 metrics. However, they provide crucial insights that complement room acoustic metrics. $L_{A,90,4h}$ (included in ISO 22955; Annex E), and a sound fluctuation parameter $M_{A,eq}$, were useful predictors of acoustic dissatisfaction and lack of privacy, respectively. The use of psychoacoustic metrics is not supported from findings here but is not ruled out either.

## SUPPLEMENTARY MATERIAL

See the supplementary material for SuppPub1, the questionnaire used, and a figure that compares $r_D$ and $L_{p,A,B}$ values across previous studies. SuppPub2 contains more detailed data about the offices sampled.

## ACKNOWLEDGMENTS

This study was funded through the Australian Research Council's Discovery Projects scheme (DP160103978). M.Y. was supported by a Deutsche Forschungsgemeinschaft (DFG) Research Grant (Project No. 503914237). The authors thank Michael Vorländer for comments on an early draft of the manuscript.

## AUTHOR DECLARATIONS
### Conflict of Interest

The authors have no conflicts of interest to disclose.

### Ethics Approval

The study protocol was approved by the University of Sydney Human Research Ethics Committee (Project No. 2017/285).

## DATA AVAILABILITY

Data available on request from the authors.

[1] R. Pirn, "Acoustical variables in open planning," J. Acoust. Soc. Am. **49**(5A), 1339–1345 (1971).

[2] S. Banbury and D. Berry, "Office noise and employee concentration: Identifying causes of disruption and potential improvements," Ergonomics **48**(1), 25–37 (2005).

[3] A. Kaarlela-Tuomaala, R. Helenius, E. Keskinen, and V. Hongisto, "Effects of acoustic environment on work in private office rooms and open-plan offices—Longitudinal study during relocation," Ergonomics **52**(11), 1423–1444 (2009).

[4] M. Pierrette, E. Parizet, P. Chevret, and J. Chatillon, "Noise effect on comfort in open-space offices: Development of an assessment questionnaire," Ergonomics **58**(1), 96–106 (2015).

[5] A. Haapakangas, V. Hongisto, M. Eerola, and T. Kuusisto, "Distraction distance and perceived disturbance by noise—An analysis of 21 open-plan offices," J. Acoust. Soc. Am. **141**(1), 127–136 (2017).

[6] S. H. Park, P. J. Lee, B. K. Lee, M. Roskams, and B. P. Haynes, "Associations between job satisfaction, job characteristics, and acoustic environment in open-plan offices," Appl. Acoust. **168**, 107425 (2020).

[7] T. Parkinson, S. Schiavon, J. Kim, and G. Betti, "Common sources of occupant dissatisfaction with workspace environments in 600 office buildings," Build. Cities **4**(1), 17–35 (2023).

[8] J. Radun and V. Hongisto, "Perceived fit of different office activities—The contribution of office type and indoor environment," J. Environ. Psychol. **89**, 102063 (2023).

[9] P. Manning, "Office Design: A Study of Environment," Department of Building Science, University of Liverpool, Liverpool, England (Thirty Shillings) (1965).

[10] M. J. Brookes and A. Kaplan, "The office environment: Space planning and affective behavior," Hum. Factors **14**(5), 373–391 (1972).

[11] A. Hedge, "The open-plan office: A systematic investigation of employee reactions to their work environment," Environ. Behavior **14**(5), 519–542 (1982).

[12] E. Sundstrom, J. P. Town, R. W. Rice, D. P. Osborn, and M. Brill, "Office noise, satisfaction, and performance," Environ. Behavior **26**(2), 195–222 (1994).

[13] J. A. Veitch, K. E. Charles, K. M. J. Farley, and G. R. Newsham, "A model of satisfaction with open-plan office conditions: COPE field findings," J. Environ. Psychol. **27**(3), 177–189 (2007).

[14] J. Kim and R. de Dear, "Workspace satisfaction: The privacy-communication trade-off in open-plan offices," J. Environ. Psychol. **36**, 18–26 (2013).

[15] L. T. Graham, T. Parkinson, and S. Schiavon, "Lessons learned from 20 years of CBE'92s occupant surveys," Build. Cities **2**(1), 166–184 (2021).

[16] N. Saval, *Cubed: A Secret History of the Workplace* (Doubleday, New York, 2014).

[17] E. Sundstrom, R. K. Herbert, and D. W. Brown, "Privacy and communication in an open-plan office: A case study," Environ. Behavior **14**(3), 379–392 (1982).

[18] F. Halldorsson, K. Kristinsson, S. Gudmundsdottir, and L. Hardardottir, "Implementing an activity-based work environment: A longitudinal view of the effects on privacy and psychological ownership," J. Environ. Psychol. **78**, 101707 (2021).

[19] V. Kupritz, "Accommodating privacy to facilitate new ways of working," J. Archit. Plann. Res. **20**(2), 122–135 (2003).

[20] ISO 22955, "Acoustics—Acoustic quality of open office spaces" (International Organization for Standardization, Geneva, Switzerland, 2021).

[21] W. J. Cavanaugh, W. R. Farrell, P. W. Hirtle, and B. G. Watters, "Speech privacy in buildings," J. Acoust. Soc. Am. **34**(4), 475–492 (1962).

[22] J. S. Bradley, "The acoustical design of conventional open plan offices," Can. Acoust. **31**(2), 23–31 (2003).

[23] V. Hongisto, "A model predicting the effect of speech of varying intelligibility on work performance," Indoor Air **15**(6), 458–468 (2005).

[24] P. Virjonen, J. Keränen, R. Helenius, J. Hakala, and O. V. Hongisto, "Speech privacy between neighboring workstations in an open office-a laboratory study," Acta Acust. Acust. **93**(5), 771–782 (2007).

[25] ISO 3382-3:2022, "Measurement of room acoustic parameters. Part 3: Open plan offices" (International Organization for Standardization, Geneva, Switzerland, 2022).

[26] P. Virjonen, J. Keränen, and V. Hongisto, "Determination of acoustical conditions in open-plan offices: Proposal for new measurement method and target values," Acta Acust. united Acust. **95**(2), 279–290 (2009).

[27] G. R. Newsham, J. A. Veitch, and K. E. Charles, "Risk factors for dissatisfaction with the indoor environment in open-plan offices: An analysis of COPE field study data," Indoor Air **18**(4), 271–282 (2008).

[28] M. Hodgson, "Acoustical evaluation of six 'green' office buildings," J. Green Build. **3**(4), 108–118 (2008).

[29] M. Yadav, D. Cabrera, J. Kim, J. Fels, and R. de Dear, "Sound in occupied open-plan offices: Objective metrics with a review of historical perspectives," Appl. Acoust. **177**, 107943 (2021).

[30] J. Nemecek and E. Grandjean, "Results of an ergonomic investigation of large-space offices," Hum. Factors **15**(2), 111–124 (1973).

[31] V. Hongisto, A. Haapakangas, J. Varjo, R. Helenius, and H. Koskela, "Refurbishment of an open-plan office—Environmental and job satisfaction," J. Environ. Psychol. **45**, 176–191 (2016).

[32] D. Cabrera, M. Yadav, and D. Protheroe, "Critical methodological assessment of the distraction distance used for evaluating room acoustic quality of open-plan offices," Appl. Acoust. **140**, 132–142 (2018).

[33] X. Wang and Z. Cheng, "Cross-sectional studies: Strengths, weaknesses, and recommendations," Chest **158**(1), S65–S71 (2020).

[34] K. D. Kryter, *The Effects of Noise on Man* (Academic Press, New York, 1970).

[35] L. Lenne, P. Chevret, and J. Marchand, "Long-term effects of the use of a sound masking system in open-plan offices: A field study," Appl. Acoust. **158**, 107049 (2020).

[36] V. Hongisto and J. Keränen, "Comfort distance—A single-number quantity describing spatial attenuation in open-plan offices," Appl. Sci. **11**(10), 4596 (2021).

[37] M. Yadav, D. Cabrera, J. Love, J. Kim, J. Holmes, H. Caldwell, and R. de Dear, "Reliability and repeatability of ISO 3382-3 metrics based on

[37] repeated acoustic measurements in open-plan offices," Appl. Acoust. 150, 138–146 (2019).

[38] H. Fastl and E. Zwicker, *Psychoacoustics: Facts and Models* (Springer, Berlin, 2007).

[39] W. Revelle, "Package 'psych,'" Comp. R. Arch. Netw. 337, 338 (2015).

[40] A. Field, J. Miles, and Z. Field, *Discovering Statistics Using R* (Sage, Thousand Oaks, CA, 2012), p. 01447.

[41] L. Hu and P. M. Bentler, "Cutoff criteria for fit indexes in covariance structure analysis: Conventional criteria versus new alternatives," Struct. Eq. Model.: Multidisc. J. 6(1), 1–55 (1999).

[42] Y. Rosseel, "lavaan: An R package for structural equation modeling," J. Stat. Softw. 48, 1–36 (2012).

[43] C. DiStefano, M. Zhu, and D. Mîndrilă, "Understanding and using factor scores: Considerations for the applied researcher," Pract. Assess. Res. Eval. 14(1), 20 (2009).

[44] P.-C. Bürkner, "brms: An R package for Bayesian multilevel models using Stan," J. Stat. Softw. 80, 1–28 (2017).

[45] J. Cohen, "A power primer," Psychol. Bull. 112(1), 155–159 (1992).

[46] P. Bottalico, "Lombard effect, ambient noise, and willingness to spend time and money in a restaurant," J. Acoust. Soc. Am. 144(3), EL209–EL214 (2018).

[47] M. Yadav and D. Cabrera, "Acoustic privacy," in *Routledge Handbook of High-Performance Workplaces* (Routledge, Milton Park, UK, 2024), pp. 102–115.

[48] S. J. Schlittmeier, J. Hellbrück, and M. Klatte, "Does irrelevant music cause an irrelevant sound effect for auditory items?," Eur. J. Cogn. Psychol. 20(2), 252–271 (2008).

[49] A. Haapakangas, E. Kankkunen, V. Hongisto, P. Virjonen, D. Oliva, and E. Keskinen, "Effects of five speech masking sounds on performance and acoustic satisfaction. Implications for open-plan offices," Acta Acust. united Acust. 97(4), 641–655 (2011).

[50] A. Haapakangas, V. Hongisto, J. Hyönä, J. Kokko, and J. Keränen, "Effects of unattended speech on performance and subjective distraction: The role of acoustic design in open-plan offices," Appl. Acoust. 86, 1–16 (2014).

[51] M. Yadav and D. Cabrera, "Two simultaneous talkers distract more than one in simulated multi-talker environments, regardless of overall sound levels typical of open-plan offices," Appl. Acoust. 148, 46–54 (2019).

[52] M. Yadav, M. Georgi, L. Leist, M. Klatte, S. J. Schlittmeier, and J. Fels, "Cognitive performance in open-plan office acoustic simulations: Effects of room acoustics and semantics but not spatial separation of sound sources," Appl. Acoust. 211, 109559 (2023).

[53] Z. Rachman, F. Aletta, and J. Kang, "Exploring soundscape assessment methods in office environments: A systematic review," Buildings 14(11), 3408 (2024).

[54] B. West, A. Deuchars, and I. Ali-MacLachlan, "Office soundscape assessment: A model of acoustic environment perception in open-plan offices," J. Acoust. Soc. Am. 156(5), 2949–2959 (2024).